\documentclass[12pt,a4paper]{article}

\usepackage{amsmath}
\usepackage{graphicx}
\usepackage{times}
\usepackage{cite}
\usepackage{ulem}
\begin{document}
\begin{titlepage}
\title{Reflective scattering at the LHC and  two--scale structure of a proton}
\author{ S.M. Troshin, N.E. Tyurin\\[1ex]
\small  \it NRC ``Kurchatov Institute''--IHEP\\
\small  \it Protvino, 142281, Russian Federation,\\
\small Sergey.Troshin@ihep.ru
}
\normalsize
\date{}
\maketitle

\begin{abstract}
   We discuss  interpretation  of the reflective scattering  mode connecting its appearance with   a resolution of a two--scale structure of a proton revealed in the  DVCS process  at Jefferson laboratory  and in the differential cross--section of elastic $pp$--scattering at the LHC at the energy of $\sqrt{s}=13$ TeV. 
\end{abstract}
\end{titlepage}
\setcounter{page}{2}
\section*{Introduction. The reflective scattering mode}

As it is well known, the  elastic scattering matrix element can be represented in a general form as the complex function:
\begin{equation}
S(s,b)=\kappa (s,b)\exp[2i\delta(s,b)]
\end{equation}
with the two real functions $\kappa$ and $\delta $ and $\kappa$ can vary in the interval
$0\leq \kappa \leq 1$. The impact parameter $b$, $b=2l/\sqrt{s}$, is a conserved quantity at high energies.  The function $\kappa$  is known as an absorption factor, its value $\kappa =0$ means a complete absorption of the initial state, it is related to the probability distribution of the inelastic interactions over impact parameter:
\begin{equation}
\kappa^2 (s,b)=1-4h_{inel}(s,b),
\end{equation}
where $h_{inel}(s,b)$ is a total contribution of the inelastic intermediate states into the unitarity relation: 
\begin{equation}
\mbox{Im} f (s,b)=|f(s,b)|^2+ h_{inel}(s,b).
\end{equation}
Note that the normalization is such that $S=1+2if$, where $f$ is the scattering amplitude. Unitarity relation in the impact parameter representation implies that the limiting behaviour $\mbox{Im}f\to 1$ leads to a vanishing real part of scattering amplitude, i.e. $\mbox{Re}f\to 0$,  cf. \cite {tr}.

It was shown in \cite{alkin1,tsrg} that  the inelastic overlap function $h_{inel}$   values  are very close to its limiting value (in the present normalisation it is
$h^{max}_{inel}=1/4$) in the rather broad region of impact parameters, i.e. till $b\simeq 0.4$ fm at the LHC energy $\sqrt{s}=13$ TeV. Deviation of $h_{inel}$ from the maximal value  is small and negative in this region of impact parameters (note, that $h_{inel}(s,b)$ has a shallow local minimum at $b=0$). Then the unitarity relation at this energy and impact parameters can be written as
\begin{equation} \label{bdh}
(\mbox{Im}f-{1}/{2})^2 +(\mbox{Re} f)^2\simeq 0.
\end{equation}
  It follows from Eq. (\ref{bdh}) that $\mbox{Re} f\simeq 0$ and $\mbox{Im} f\simeq 1/2$. Thus, the observed impact parameter picture of elastic scattering together with unitarity  can provide at least a qualitative explanation of the recent  result on the unexpectedly small  real to imaginary parts ratio of the forward  scattering amplitude \cite{ro}.  Indeed,  the above region of impact
 parameters being a most significant one provides leading contributions to the real and imaginary parts of a forward scattering amplitude\footnote{Those parts are the respective integrals  of the functions $\mbox{Re} f$ and $\mbox{Im} f$ over impact parameter.}. There is no room for a significant value of the real part of elastic scattering amplitude at $\sqrt{s}=13$ TeV.
Therefore, the real part of the elastic scattering amplitude $f$ can be safely  neglected due to its smallness (cf. for the numerical estimations \cite{drem}, qualitative arguments have been given above). And the following simplifying replacement will be used: $f\to i f$.

 The function $S(s,b)$ can be a nonnegative one in the whole region of the impact parameter variation or it can acquire   negative values in the region $b<r(s)$ at high enough energy, i.e. at $s>s_r$, where $s_r$ is  a solution of the equation $S(s,b=0)=0$ (note that $r(s)$ is defined as $S(s,b=r(s))=0$ at $s>s_r$)  The $s$--dependent function $r(s)$ increases as $\ln s$ at $s\to\infty$\cite{ijmpa07} and its value at $\sqrt{s}=13$ TeV is about $0.4$ fm. This is the region of energies where the reflection appears, i.e the function $S(s,b)$ crosses zero at $b=r(s)$ and  the value of $\delta$ jumps from $0$ to 
 $\delta = \pi/2$ at this point. Since reflective scattering is not a commonly accepted nomenclature  nowadays,  a brief reminder of the main features of this mode is necessary.

Under the reflective scattering, $f>1/2$, an increase of elastic scattering amplitude $f$ correlates with decrease of $h_{inel}$ according to unitarity relation
\begin{equation}
(f-{1}/{2})^2={1}/{4}-h_{inel}=\kappa^2 /4
\end{equation}
and the term antishadowing has initially been  used \cite{asd}  emphasizing that the reflective scattering is correlated with the self-damping of the inelastic channels contribution \cite{baker} and   decoupling of the elastic scattering from multiparticle production dynamics with energy increase. This phenomena is referred nowadays as hollowness \cite{lsl, bron1, soto, soto1}.

The negative values of $S(s,b)$ correspond to the  value of $\delta=\pi/2$.  The term reflective has been borrowed from optics
where phases of incoming and outgoing  waves differ by $\pi$. Such phase jump takes place when the reflecting medium gets higher optical density (i.e. it has  a higher refractive index) than the  medium where incoming wave comes from. Optical density is then energy--dependent function.
Here there is an analogy with the sign change   of the electromagnetic wave under its reflection by  the surface of a  conductor. 
The energy evolution of the effective scatterer   leads to appearance of  the reflective scattering mode (provided the unitarity saturation takes place in the limit of $s\to\infty$). 

Reflective scattering mode does not imply
any kind of hadron transparency in the head-on collisions. Rather, it is about the geometrical elasticity \cite{usprd}.  The term transparency is relevant for the energy and impact parameter range related to the shadow scattering regime only, i.e. where $f<1/2$. The interpretation of the reflective scattering mode which is based   on the consideration of inelastic overlap function alone is, therefore, a deficient one, it does not provide any information on the collision elasticity.

The emerging physical picture of   high energy  hadron interaction region  in transverse plane  can be visualized  then in the form of
 a reflective
disk (with its albedo approaching to complete reflection at the center) which is surrounded by a   black ring 
(with complete absorption, $h_{inel}=1/4$) since the inelastic overlap function $h_{inel}$ has a prominent peripheral form  at $s\to\infty$ in this scattering mode.
The reflection  mode implies that the following  limiting behavior\footnote{Despite the limiting behavior of $S(s,b)$ corresponds to $S\to -1$ at $s\to\infty$ and fixed $b$, the gap survival probability tends to zero at $s\to\infty$  and, in particular, contribution of the central  production processes is consistent with unitarity  contrary to conclusion of  \cite{khoze}, see for details \cite{gsp}.}  
\begin{equation}
\label{neg}
S(s,b)|_{b=0}\to -1
\end{equation}
 will take place at $s\to\infty$ due to  self--damping of the contributions of the inelastic channels \cite{baker}.  Asymptotic growth of the total cross--section corresponds to saturation of unitarity and comes from an energy increase of the reflective disk radius and its albedo. Inelastic processes give a subleading contribution at $s\to\infty$.

Of course, it is considered that a monotonic increase  of the amplitude $f$ with energy to its unitarity limit $f=1$ takes place, and an unrealistic option  of its nonmonotonic energy dependence at fixed values of $b$  is excluded.

QCD is a  theory of hadron interactions  with colored objects confined inside those entities.   One can imagine that the color conducting medium is being formed instead of color insulating one when the energy of the colliding hadrons increases beyond some threshold value. Properties of such a medium are under active studies in nuclear collisions, but color conducting phase can be generated in hadron interactions too.   Appearance of the reflective scattering mode can be associated  with formation of the color conducting medium in the intermediate state of hadron interaction \cite{jpg19}.

The two recent experiments at JLab and the LHC  \cite{burk,dstot}   have proved to be significant and complementary for understanding the proton structure as well as the structure of the proton interactions region. Usage of the impact parameter picture and address to the reflective scattering  mode  allow combining  analysis of the results from those experiments for  physics interpretation of the reflective scattering.

 \section{Picture of  a proton in soft processes}
 One can address the problem of the  microscopic interpretation of the reflective scattering on the base of the new observations of the inner proton structure obtained in Jefferson Laboratory \cite{burk}. Those are in favor of the hypothesis of the dominance of elastic scattering in the deconfined mode.
 
 An existence of a strong positive repulsive pressure has been detected at the center of the proton under investigations of its structure.
 The value of this repulsive pressure exceeds the one in the neutron stars \cite{burk}. The binding (negative)  pressure exists in the peripheral part of the proton. One should note that such pressure distribution is a typical one for the chiral quark-soliton model \cite{goeke, waka} where constituent quarks are confined due to interaction with a self-consistent pion field. Similar pressure distribution can also be expected in the models for hadron structure and their interactions proposed in \cite{islam,jenk,gls,chi,pump}. 
 
 In general, the soft hadron interactions are described by  the nonperturbative sector of QCD. In this regime QCD should provide the two important phenomena: confinement (scale $\Lambda_{QCD}=100-300$ MeV) and spontaneous breaking of chiral symmetry ( $\Lambda_{\chi}\simeq 4\pi f_\pi\simeq 1$ GeV). Chiral symmetry is spontaneously broken between these two scales and this breaking generates quark masses. Since the soft hadron interactions occur at the  distances where chiral symmetry is spontaneously broken one can conclude that a major role in such interactions belongs to constituent quarks \cite{royzen} which are the colored but  not  pointlike objects. In addition to acquiring the masses, the strong interaction dynamics provides them with finite sizes, too. Chiral models describe baryon as consisting from an inner core with baryonic charge and an outer cloud surrounding the core.  Interpretation of the results of the CLAS experiment at Jefferson Laboratory based on the existence of the extended substructures inside the proton was proposed  in \cite{petr}.  Presence of the inner repulsive core is in agreement with the recent direct DVCS data \cite{burk} and with the indirect LHC data at $\sqrt{s}=13$ TeV \cite{dstot,tsrg}.
 
 On the above basis one can assume that the different aspects of hadron dynamics are represented in the following form of the effective Lagrangian  \cite{gold}:
 \begin{equation}\label{lagr}
{ \cal {L}}_{eff}=\cal {L}_\chi+\cal {L}_I+\cal {L}_C,
  \end{equation}
  where term $\cal {L}_\chi$ is  responsible for providing constituent quarks the finite masses and sizes,
 $ \cal {L}_I$ desribes their interaction mediated by the Goldstone bosons and $\cal {L}_C$ --- the color confinement. The latter term is switched off when the reflective scattering mode appears,  it is switched off first in the central collisions (deconfinement) like it happens in the bag model \cite{low}) and in this mode the geometric  elastic scattering of hadron cores starts to appear and becomes noticeable.  Thus, confining pressure begins to disappear in the high--energy central collision at $s>s_r$ and changes in the particle production mechanism would take place: maximal probability  of the secondary particles production will takes place in the peripheral collisions, $b\neq 0$.

 \section{Second  diffraction cone in $d\sigma/dt$  as a result of the  proton's cores interaction.}
 The second exponential slope \cite{dstot,tsrg} observed in the differential cross-section behavior at the LHC at large values of $-t$  is in favor of the presence of a core in the hadron structure and color deconfinement under collisions with small values of $b$. Indeed, due to deconfinement the scattering in the  deep-elastic region becomes sensitive to   presence of the inner core. Thus, one can consider  the two exponential slopes observed in the differential cross-section at the LHC as a consequence of the two--component structure of a proton.           
 The idea of  a proton  core  is not  new at all, it has been discussed by Orear in \cite{orear}.  A core is a typical feature of various chiral models representing  baryon as an inner core carrying baryonic charge and an outer cloud  \cite{heines}.

 The outer cloud of the proton is responsible for confinement. The interactions of the proton's clouds are responsible for   multiparticle  reactions \cite{chern}. Those interactions lead to   the first exponential cone observed at small  transferred momenta in the elastic processes. This first cone appears as a result of unitarity relation  connecting elastic and inelastic scatterings.
 
 Note, however, that Orear type dependence of $d\sigma/dt$ in the region of $-t$ beyond the dip,  
 \begin{equation}\label{or}
 d\sigma/dt\sim\exp({-\tilde{b}_2\sqrt{-t}}),
 \end{equation}
  is in a good agreement with  the data at moderate energies  (CERN ISR data) in a rather wide range of $-t$ variation \cite{ech}. This dependence is considered usually as a result of  a contribution of the branch points in the complex angular momentum plane generated by
  multiple rescatterings   as a direct result of unitarity  in the shadow scattering region where applicable \cite{ans,anddr,sav}. 
 
 It was shown in \cite{7tev} that use of the functional dependence (\ref{or}) at the LHC ($\sqrt{s}=7$ TeV)  fits the experimental data better than the power-like dependence and it was claimed that utilization  of the power--like dependence is not preferable since a rather  limited range of transferred  momenta (till the value of $-t=2.5$ (GeV/c)$^2$) has been covered at  $\sqrt{s}=7$ TeV.  However, such extension to the higher LHC energies in its turn has appeared to be again a not preferable one also due to a too narrow range (but not due to the collision energy value) of transferred momenta covered.
 
 Currently, the data are available for the energy $\sqrt{s}=13$ TeV \cite{dstot}. Increase of the collision energy has allowed one to expand  the range of transferred momenta available for the measurements up to $-t\simeq 4$ (GeV/c)$^2$. It has appeared now that exponential dependence on ${-t}$  fits the data significantly better than exponential dependence of $\sqrt{-t}$. Of course, this conclusion should be taken with caution since  the range of available  transferred momenta is  not too wide again.   At the moment we have no experimental data for $d\sigma/dt$  at the LHC energies in the range of $-t$ which was available for the measurements  at the CERN ISR $\sqrt {s}=53$ GeV. However,  a good agreement of Philips--Barger parameterization of amplitude with the experimental data on $d\sigma/dt$ (cf. \cite{bras} and references therein) is in favor of the above conclusion on presence of the second cone in the differential cross--section of elastic scattering. The slope parameter of the second cone is an energy--dependent one. It has a similar energy--dependence to the slope parameter  of the first cone. This similarity, in particular, is implied by the scaling dependence of the rescaled differential cross--section of elastic $pp$--scattering at the LHC energies. This scaling has been discussed in  \cite{tsrg1}.
 
  We  suggest that the observed deviation of $d\sigma/dt$ from the Orear dependence  \begin{equation}  \exp({-\tilde{b}_2\sqrt{-t}})\end{equation} at the LHC is in favor of the functional  dependence of $d\sigma/dt$ in the form of a linear exponent  \begin{equation} \exp({-b_2t})\end{equation} in the region of transferred momenta beyond the dip  is a manifestation of a second scale in the hadron structure which in its turn is related to the reflective scattering mode. The magnitude of the ratio of the two scales sizes is correlated with the ratio of the first and second slopes values of $d\sigma/dt$ dependence on $-t$. 
  These ratios can be approximately connected by the  relation:
 \begin{equation}
 b_1/b_2\simeq (r_p/r_c)^2,
 \end{equation}
 where $r_p$ is the proton radius and $r_c$ is the radius of its core.  From the experimental data at $\sqrt{s}=13$ TeV one can approximate the value of a core radius as $$r_c\simeq 0.5r_p$$ (cf. e.g. \cite{dremn}).
 
 Thus, the proton in soft processes resembles a hard  ball coated with a  thick but fragile shell. Energy evolution of the proton interactions can be imagined as follows.  Interactions of the fragile shells (responsible for the inelastic processes)  dominates up to the LHC region of energies and elastic scattering at those energies is just a shadow of  the inelastic processes, while at the LHC  the elastic scattering becomes sensitive to the core interactions in the central collisions (at $b\simeq 0$) first and  acquires  a pure geometrical component \cite{geom} due to high energy of collision.  Interaction of the cores provides thus, in particular, a second exponential cone in $d\sigma/dt$  and the elastic interaction gradually acquires a geometric nature at small values of the impact parameter. Radius of the core interactions is correlated with the solution of equation 
 \begin{equation}
 h_{el}(s,b)=1/4,
 \end{equation}
at fixed energy $s>s_r$, where $h_{el}(s,b)=f^2(s,b)$ is the elastic overlap function. This solution is determined by  dynamics of the peripheral clouds and the cores interaction. 

As a result an increase of the ratio $\sigma_{el}(s)/\sigma_{tot}(s)$ \cite{19} takes place due to  redistribution of probabilities between elastic and inelastic interactions in favor of the elastic ones. This redistribution starts to appear in the region of small impact parameters  and  proliferates into the region of higher impact parameter values at higher energies. At the LHC energy $\sqrt{s}=13$ TeV this region cover the range of impact parameters $0\leq b\leq 0.4$ fm.
 The mechanism will lead to the  elastic scattering dominance at $s \to \infty$, increasing with energy decoupling of elastic scattering from inelastic production proceses \cite{del} and  respective slow down of the mean multiplicity growth at the LHC energies and beyond \cite{sld}. It gradually turning into a driver of the total cross-section growth  at  high energies.
  \begin{figure}[hbt]
 	\vspace{-0.5cm}
 	\hspace{-1cm}
 	\resizebox{16cm}{!}{\includegraphics{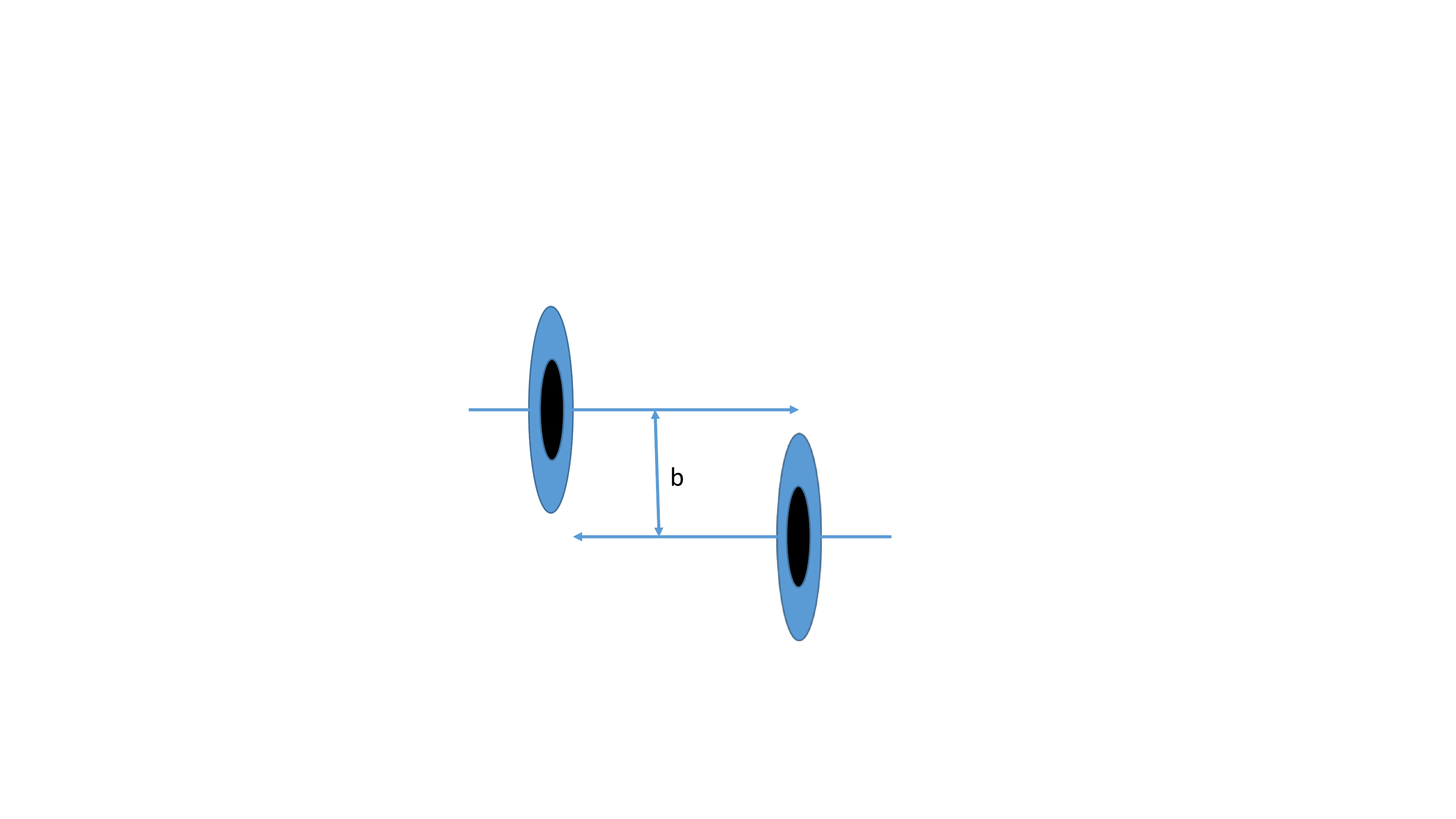}}		
 	\vspace{-2cm}
 	\caption{Two--components' protons scattering  at the  impact parameter $b$.}	
 \end{figure}	 
   \begin{figure}[hbt]
  	\vspace{-0.5cm}
  	\resizebox{16cm}{!}{\includegraphics{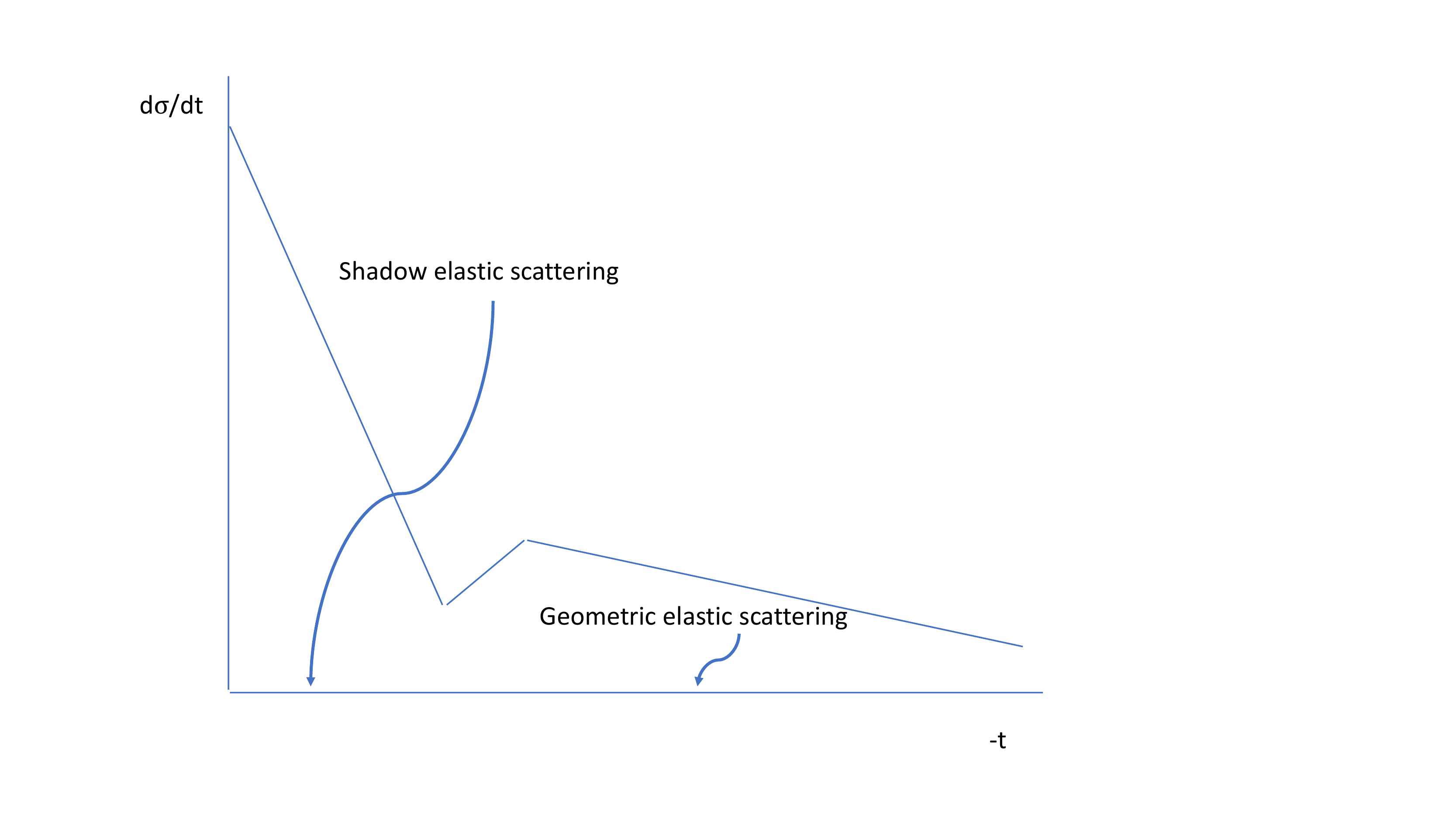}}		
  	\vspace{-1cm}
  	\caption{Two regions of transferred momenta $-t$  (relevant for shadow and geometric elastic scattering) in  $d\sigma/dt$  at the LHC energies. The size of  shadow scattering region diminishes  with  energy (decoupling of elastic scattering from multiparticle production) and tends to zero at $s\to\infty$.}	
  \end{figure}	 
 
  The form and size of the  interaction region with reflective (geometric) scattering contribution in the impact parameter space are consistent with the  impact parameter analyses at the LHC energy $\sqrt{s}=13$ TeV \cite{alkin1,tsrg}. As it was shown in \cite{tsrg}, a statistical significance of the ``black ring'' effect at this energy is greater than $5\sigma$ , the real part of the scattering amplitude  gives a very small contribution  as it was expected and does not change the result. Thus, the existence of the ``black ring'' effect should be considered now as an experimentally established fact. As it was noted in \cite{tsrg} a dip at $b=0$ becomes a generic property of the inelastic overlap function at high enough energies. 
  
  The appearance of the reflective scattering mode can be interpreted as a result of the color conducting phase formation at high energies. Therefore,  presence of reflective scattering mode contribution emphasizes  importance of the events classification according to the  impact parameter of the collision since the reflective scattering affects  those with small impact parameters \cite{cent}. 
  
  Thus, one can qualitatively conclude  that  behavior of the differential cross--section at the LHC energies  follows a linear exponential dependence in  both regions corresponding to the shadow scattering and to geometric scattering. Region of transferred momenta where the geometric  scattering\footnote{Geometrical  scattering gives, however, a rather small contribution to the integrated cross--section of elastic scattering $\sigma_{el}(s)$ at the LHC energies.} is dominating increases with energy and results in shift of the dip in the differential cross--section $d\sigma/dt$  into the region of smaller values of $-t$. The two regions  are schematically represented at Fig. 2. Asymptotically, domination of geometric scattering would lead to saturation of unitarity limit, i.e. to a flat form of a scattering amplitude at small and moderate impact parameters which would results in typical diffraction pattern of the differential cross-section with many secondary maxima and minima \cite{asp} similar to the one observed in nuclei collisions \cite{ble}.

 It should  be emphasized that the collision geometry describes   the  hadron interaction region   but not  the spacial properties of the individual participating hadrons. This geometry is determined by both a structure of interacting hadrons and   dynamics of their interactions.

 \section{Conclusions}
  
  The general statement  is that an expected transition from the shadow to geometric elastic scattering could start to occur already at the LHC energies. The claim is supported, in particular, by the interpretation of the second cone observed at the LHC as a consequence of a core   in the hadron structure. 
 
 Dynamics of elastic $pp$--scattering  is described by a  function $F(s,t)$ (spin degrees of freedom of the colliding protons are neglected) of the two Mandelstam variables $s$ and $t$, the latter variable is a conjugated  to the  impact parameter $b$. 
 As it was  noted,  any quantity integrated over $b$   is not sensitive to details of its dependencie  and therefore it cannot provide an sufficient information  for  conclusions on the interaction dynamics. 
  
On the base of $b$-dependent consideration of elastic and inelastic interactions there was proposed a  connection of the reflective scattering mode with formation of color--conducting medium in the intermediate state at high energies and small  impact parameter values \cite{jpg19}. This analogy is based  on replacement of an electromagnetic field  by a chromomagnetic field of QCD.   One can also address the phenomenon of Andreev reflection  at the boundary of the normal and superconducting phase \cite{and, sadz}. 

Presence of  the core in  a proton can be considered as a result of the different values of the scales $\Lambda_{QCD}$ and  $\Lambda_{\chi}$  relevant for the two important phenomena in the nonperturbative QCD: confinement and spontaneous chiral symmetry breaking. The latter phenomena seems to be relevant for a core formation.  As it was noted before,  the proton structure can be imagined as a hard  ball placed in the hadron central region and  coated by a fragile peripheral stuff. This hypotheses on core in a proton is relevant for the soft interactions, we do not concern here the hard ones where hadron's structure at the very short distance is important and where the parton model of hadrons with perturbative QCD are working well. The problem of transition beetwen these two pictures is correlated with  the problem of transition from ${ \cal {L}}_{QCD}$ to ${ \cal {L}}_{eff}$.  Above structure   devoid an 
assumption\footnote{This assumption corresponds to the black limit saturation at $s\to\infty$, cf. e.g. \cite{wu},  with equipartition of the elastic and inelastic cross--sections in this limit.} of particle production  the status of a leading driver of an asymptotic hadron interaction dynamics. The emergent substitution, i.e. hypothesis of maximal importance of elastic scattering, is based on the  saturation of unitarity due to a maximal strength of strong interactions (i.e. maximality of the imaginary part of the partial amplitude of elastic scattering consistent with unitarity constraint). This principle has been developed by Chew and Frautchi   along with the ``strip approximation'' \cite{ch}.

No doubt, future experimental measurements  would be crucial for the studies of hadron dynamics in the nonperturbative sector of QCD and allow one, in particular, to perform a sensible   choice among shadow and geometric elastic scattering hypotheses, i.e. between the principles of maximal importance of particle production and  maximal strength of strong interactions  in the limit $s\to\infty$.

\section*{Acknowledgements}
We are grateful to T. Cs\"{o}rg\H{o} for the interesting and useful correspondence on the two--scale structure of a proton in the context of the recent data of the LHC on the differential cross--section of elastic scattering at $\sqrt{s}= 13$ TeV.

\end{document}